\documentclass[aps,prd,showpacs,twocolumn,nofootinbib,superscriptaddress]{revtex4}

\pdfoutput=1
\usepackage{cancel}
\usepackage{amssymb,amsfonts,amsmath}
\usepackage{graphicx,epsfig,pdfpages,hyperref}
\usepackage{caption}
\usepackage{subcaption}
\newcommand{\bea}{\begin{eqnarray}}
\newcommand{\eea}{\end{eqnarray}}

\def\gsim{\mathrel{
   \rlap{\raise 0.511ex \hbox{$>$}}{\lower 0.511ex \hbox{$\sim$}}}}
\def\lsim{\mathrel{
   \rlap{\raise 0.511ex \hbox{$<$}}{\lower 0.511ex \hbox{$\sim$}}}}

\def \chonep{{\wt\chi_1}^{+}}
\def \chonem{{\wt\chi_1^-}}
\def \chonep2{{\wt\chi_2^+}}
\def \chonem2{{\wt\chi_2^-}}

\def\issue(#1,#2,#3){{\bf #1}, #2 (#3)}
\def\PREP(#1,#2,#3){Phys.\ Rep. \issue(#1,#2,#3)}

\hypersetup{
   bookmarks=true,         
   unicode=false,          
   pdftoolbar=true,        
   pdfmenubar=true,        
   pdffitwindow=false,     
   pdfstartview={FitH},    
   pdfauthor={Author},     
   pdfsubject={Subject},   
   pdfcreator={Creator},   
   pdfproducer={Producer}, 
   pdfkeywords={keyword1} {key2} {key3}, 
   pdfnewwindow=true,      
   colorlinks=true,       
   linkcolor=blue,        
   citecolor=red,         
   filecolor=magenta,      
   urlcolor=cyan,           
   linktocpage = true,
   }

\begin{document}
\begin{flushleft}
\hspace*{0.5cm}{TIFR/TH/15-35}
\end{flushleft}

\title{Probing $(g-2)_{\mu}$ at the LHC in the paradigm of R-parity
violating MSSM}

\preprint{TIFR/TH/15-35}

\author{Amit Chakraborty}
\email{amit@theory.tifr.res.in}
\author{Sabyasachi Chakraborty}
\email{sabya@theory.tifr.res.in}
\affiliation{Department of Theoretical Physics, Tata Institute of Fundamental 
Research, \\ 1, Homi Bhabha Road, Mumbai 400005, India.}

\begin{abstract}
The measurement of the anomalous magnetic moment of the muon exhibits a long standing
discrepancy compared to the Standard model prediction. In this paper, we concentrate
on this issue in the framework of $R$-parity violating Minimal supersymmetric standard
model.
Such a scenario provides substantial contribution to the anomalous magnetic moment of 
the muon while satisfying constraints from low energy experimental observables as well 
as neutrino mass. In addition, we point out that the implication of such operators 
satisfying muon $g-2$ are immense from the perspective of the LHC experiment, leading 
to a spectacular four muon final state. We propose an analysis in this particular channel 
which might help to settle the debate of R-parity violation as a probable explanation 
for $(g-2)_{\mu}$.
\end{abstract}
\maketitle
\section{Introduction}
\label{intro}
We are living in an era enriched with many experimental breakthroughs and results.
Recently, the two CERN based experiments, namely ATLAS and CMS collaborations 
have confirmed the existence of a neutral boson, widely accepted to be the Higgs boson 
with mass close to 125 GeV \cite{higgs}. All the decay modes of this scalar boson have 
been measured with moderate accuracy and the results obtained so far are fairly consistent 
with the standard model (SM) expectation. However, from an aesthetic point of view, 
the SM inevitably has the hierarchy problem which is associated to the stabilization 
of the Higgs boson mass from large radiative corrections. Further, the observation of 
neutrino mass and mixing and the existence of dark matter (DM) most certainly require 
beyond the standard model (BSM) physics. Another sector which requires the intervention 
of BSM theories is the anomalous magnetic moment of the muon, quantified as $a_\mu=
(g-2)_\mu/2$, which has been measured with unprecedented accuracy at the Brookhaven
$(g-2)$ experiment. However, there still exists a discrepancy between the experimental 
observation and the SM prediction, given by $\Delta a_{\mu}=a_{\mu}^{\text{exp}}-
a_{\mu}^{\text{SM}}=(29.3\pm 9.0)\times 10^{-10}$ \cite{g-2-Collab}. This anomaly with 
respect to the SM expectation reflects the contributions arising from the perturbative 
higher order electroweak corrections, the virtual hadronic inputs and the possible 
presence of the BSM physics. 

Supersymmetry (SUSY) \cite{SUSYreviews1,SUSYreviews2,SUSYbooks,djouadimssmrev} 
remains one of the most celebrated BSM theories till date. The minimal supersymmetric 
standard model (MSSM) provides an elegant solution to the hierarchy problem 
\cite{SUSYreviews2,djouadimssmrev}. In addition neutrino masses and DM can also be 
explained in the paradigm of MSSM. Another important feature of MSSM is that it yields 
sizeable contribution to the muon $(g-2)$ requiring light first two generation of 
sleptons \cite{g-2-review1,g-2-review2}. However, the ATLAS and CMS experiments, in their
hunt for superpartners, have found no significant excess over the SM background 
after the 7+8 TeV run of the LHC \cite{Aad:2015iea,Khachatryan:2015lwa}. For comparable 
gluino and first two generation squark masses, the bound on these particles can 
be as large as 1.7 TeV in $R$-parity conserving (RPC) and simplified phenomenological 
MSSM (pMSSM) scenario \cite{LHCupdate}. On the other hand, the constraints on the
first two generation sleptons are comparatively weaker and lies in the ballpark of
300 GeV \cite{Aad:2014vma,Khachatryan:2014qwa}

In MSSM, the loop contributions to $(g-2)_{\mu}$ arises if there is a chirality 
transition in the external muon lines. This chirality transition requires an insertion 
of a fermion mass or a Yukawa coupling vertex. In the framework of $R$-parity conserving 
SUSY, the main possibilities for the chirality flip are the following, a) a muon 
line through a muon mass term, which contributes to a factor $m_{\mu}$, b) a Yukawa 
coupling in between the Higgs field and $\mu_L$, $\mu_R$ which contributes to a factor 
$y_{\mu}$, c) a $L-R$ mixing in the scalar sector, more precisely corresponding to a 
transition between $\widetilde\mu_L$-$\widetilde\mu_R$, which contributes to a factor 
proportional to $m_{\mu}\mu\tan\beta$, where $\mu$ is the Higgsino mass parameter and 
$\tan\beta$ is the ratio of two vacuum expectation values (vevs) $v_u$ and $v_d$ associated 
with the two Higgs doublets $H_u$ and $H_d$ respectively. Finally d) a SUSY Yukawa 
coupling of a Higgsino to muon and $\widetilde\mu$ or $\widetilde\nu_\mu$, contributing a 
factor of $y_\mu$. It is evident that all of these contribute to 
the muon $(g-2)$ and an overall rough estimate implies $a_\mu \sim m_{\mu}^2/
M_{\text{SUSY}}^2$~\cite{Stockinger:2006zn}. Therefore, the new physics scale or more 
precisely the SUSY scale must be around $\mathcal O(100)$ GeV, i.e., the electro-weak 
scale to have large contributions to $(g-2)_{\mu}$. 

On the other hand, one of the many interesting outcomes of $R$-parity violating
(RPV) MSSM \cite{rpvreview} is that it is an intrinsic way by which substantial 
augmentation of muon $(g-2)$ can be obtained~\cite{Kim:2001se}.
Further RPV is also interesting as it is an inherent SUSY
way to generate neutrino masses both at the tree level as well as at the one loop
level. In this work, we consider a RPV MSSM scenario with relevant operators, 
which can give sizeable contribution to the 
anomalous magnetic moment of the muons. We respect the collider bounds on the
slepton masses as well as indirect constraints from neutrino masses and low energy
observables to present a self consistent picture. 
Most importantly, RPV MSSM also provide direct spectacular signals at the LHC. 
It is important to note that we ignore
the contributions originating from left scalar and right scalar fermion mixing terms
as they are negligible. There exists several phenomenological studies 
incorporating the muon $(g-2)$ anomaly and the LHC bounds in R-parity conserving 
and violating SUSY framework, a partial list can be seen in 
Refs.~\cite{Chattopadhyay:2001vx,Chattopadhyay:1995ae,Chattopadhyay:2000ws,
Ghosh:2012ag,Chakraborti:2014gea,
Bhattacherjee:2013tha,Chakrabortty:2015ika,Bhattacharyya:2013xma,Bhattacharyya:2009hb,
Padley:2015uma}.

The plan of this paper is as follows. In section II, we look into the theoretical
framework of the study under consideration and its effects on $(g-2)_{\mu}$. 
In section III, we study the relevant constraints coming from low
energy observables and neutrino masses, which is necessary for considering a 
$\mathcal O(1)$ RPV coupling. Section IV is dedicated to a numerical analysis
with appropriate benchmark points followed by a detailed discussion on the 
present bounds from LHC data. In section V, we perform a dedicated collider analysis 
to correlate the fact that $(g-2)_{\mu}$ from the RPV MSSM scenario can leave its 
finger prints in the LHC experiments. Concluding remarks and 
related discussions are relegated to section VI.
\section{Muon $(g-2)$ in MSSM}
\label{framework}
When $R$-parity is conserved, the SUSY effects on $a_{\mu}$ includes contribution
from the chargino-muon sneutrino and neutralino-smuon loops. The
generic expressions for one-loop SUSY contributions to $a_{\mu}$, including
the effects of possible complex phases are given as \cite{Martin:2001st,Moroi:1995yh}
\begin{widetext}
\begin{eqnarray}
a_{\mu}^{\widetilde\chi^0}&=&\frac{m_{\mu}}{16\pi^2}\sum_{i,m}\Big\{
-\frac{m_{\mu}}{12 m^2_{\widetilde\mu_m}}(|n^L_{im}|^2+|n^R_{im}|^2)F_1^N(x_{im}) 
+\frac{m_{\widetilde\chi_i^0}}{3m^2_{\widetilde\mu_m}}\text{Re}(n^L_{im}n^R_{im})F_2^N(x_{im})\Big\},
\label{eq:a1}
\end{eqnarray}
\begin{eqnarray}
a_{\mu}^{\widetilde\chi^+}&=&\frac{m_{\mu}}{16\pi^2}\sum_{k}\Big\{
\frac{m_{\mu}}{12 m^2_{\widetilde\nu_{\mu}}}(|c_k^L|^2+|c_k^R|^2)F_1^C(x_k)+
\frac{2m_{\widetilde\chi_k^{\pm}}}{3m^2_{\widetilde\nu_{\mu}}}\text{Re}[c_k^Lc_k^R]F_2^C(x_k)
\Big\}
\label{eq:a2}
\end{eqnarray}
\end{widetext}
where $i=1,2,3,4$, $m=1,2$ and $k=1,2$ denotes the neutralino, smuon and chargino mass eigenstates
respectively. The couplings are defined as
\begin{eqnarray}
n^R_{im}&=& \sqrt{2}g_1 N_{i1}X_{m2}+y_{\mu}N_{i3}X_{m1}, \nonumber \\
n^L_{im}&=& \frac{1}{\sqrt{2}}(g_2 N_{i2}+g_1 N_{i1})X^*_{m1}-y_{\mu}N_{i3}X^*_{m2}, \nonumber \\
c^R_k&=& y_{\mu}U_{k2}, \nonumber \\
c^L_k&=&-g_2 V_{k1},
\end{eqnarray}
{where $N$ represents neutralino, $U$ and $V$ are chargino mixing matrices 
respectively while $X$ denotes the slepton mixing matrix.} 
The muon yukawa coupling $y_{\mu}=g_2 m_{\mu}/\sqrt{2} m_W \cos\beta$ and the kinematic loop
functions are defined in terms of the variables $x_{im}=m^2_{\widetilde\chi_i^0}/m^2_{\widetilde\mu_m}$ and
$x_k=m^2_{\widetilde\chi_k^{\pm}}/m^2_{\widetilde\nu_{\mu}}$ and are as follows
\begin{eqnarray}
F_1^N(x)&=&\frac{2}{(1-x)^4}\Big[1-6x+3x^2+2x^3-6x^2 \ln x\Big], \nonumber \\
F_2^(x) &=&\frac{3}{(1-x)^3}\Big[1-x^2+2x\ln x\Big],\nonumber \\
F_1^C(x)&=&\frac{2}{(1-x)^4}\Big[2+3x-6x^2+x^3+6x\ln x\Big],\nonumber \\
F_2^C(x)&=&-\frac{3}{2(1-x)^3}\Big[3-4x+x^2+2\ln x\Big].
\end{eqnarray}
In the limit when all the mass scales are roughly of the same order, i.e., $M_{\text{SUSY}}$, 
the sum of the above expressions in Eq.~\ref{eq:a1} and Eq.~\ref{eq:a2} reduces to a more 
simpler form as \cite{Martin:2001st,Moroi:1995yh} ({see Appendix for more detail})
\begin{eqnarray}
\Big[a_{\mu}^{\text{SUSY}}\Big]_{\text{RPC}}&\simeq& 14\tan\beta \Big(\frac{100\text{GeV}}
{M_{\text{SUSY}}}\Big)^2 10^{-10}.
\label{ref-apen}
\end{eqnarray}

\begin{figure}%
    \centering
    {{\includegraphics[width=6cm]
    {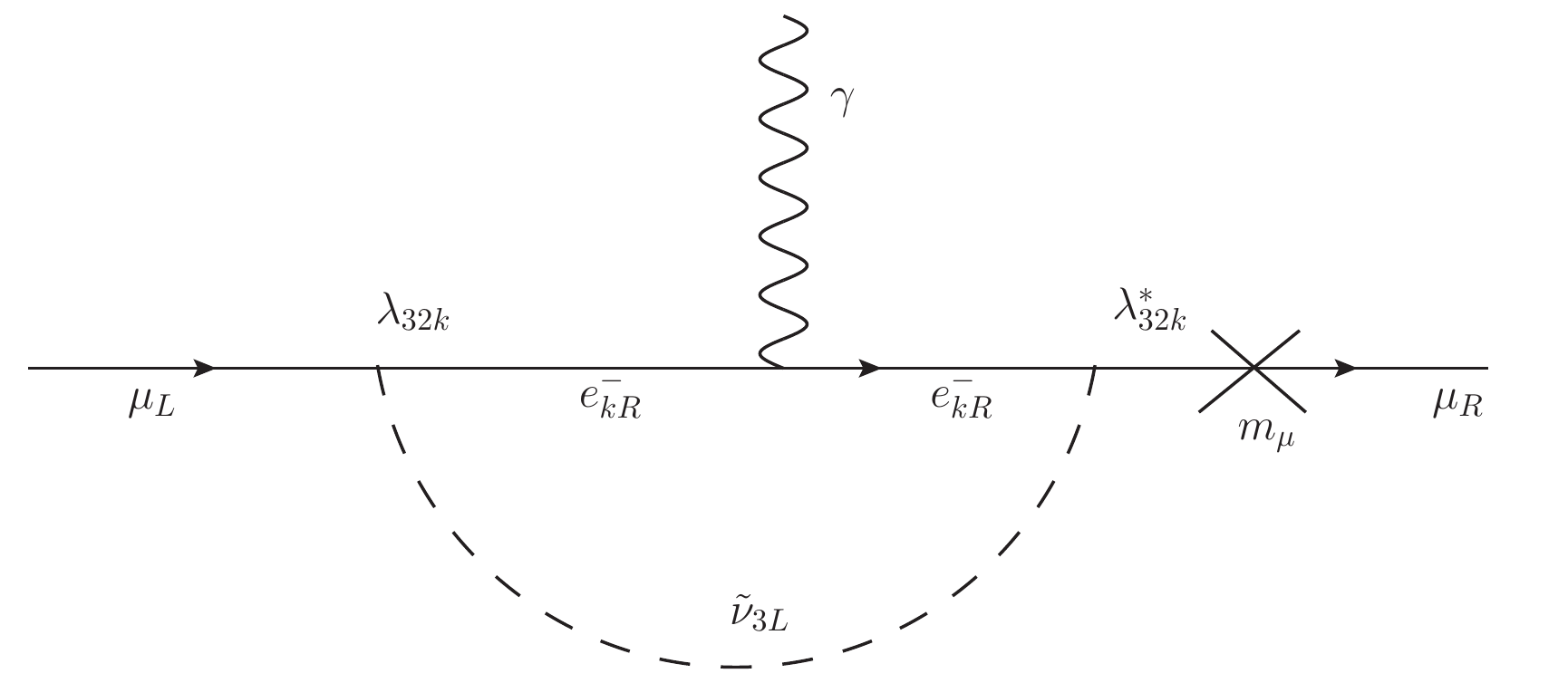} }}%
    \qquad
    {{\includegraphics[width=6cm]
    {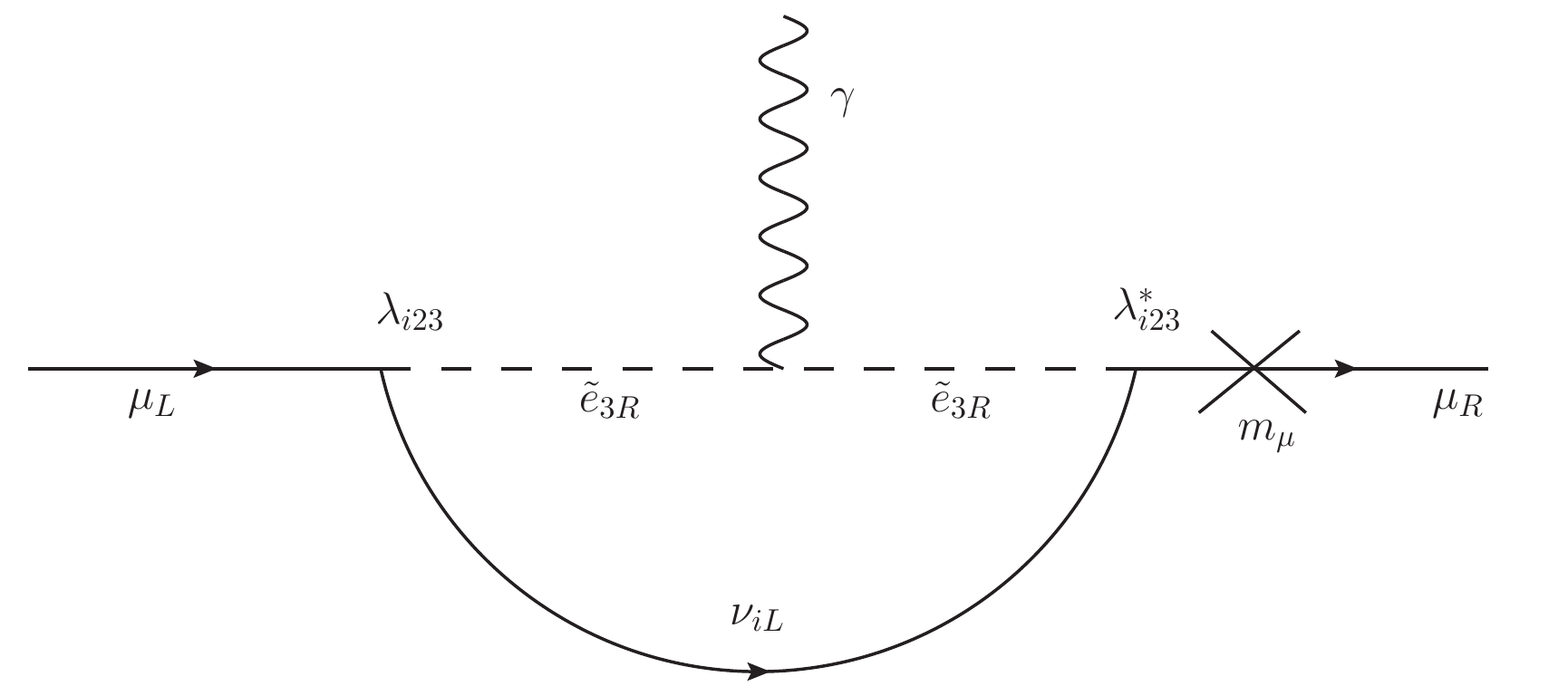} }}%
    \qquad
    {{\includegraphics[width=6cm]
    {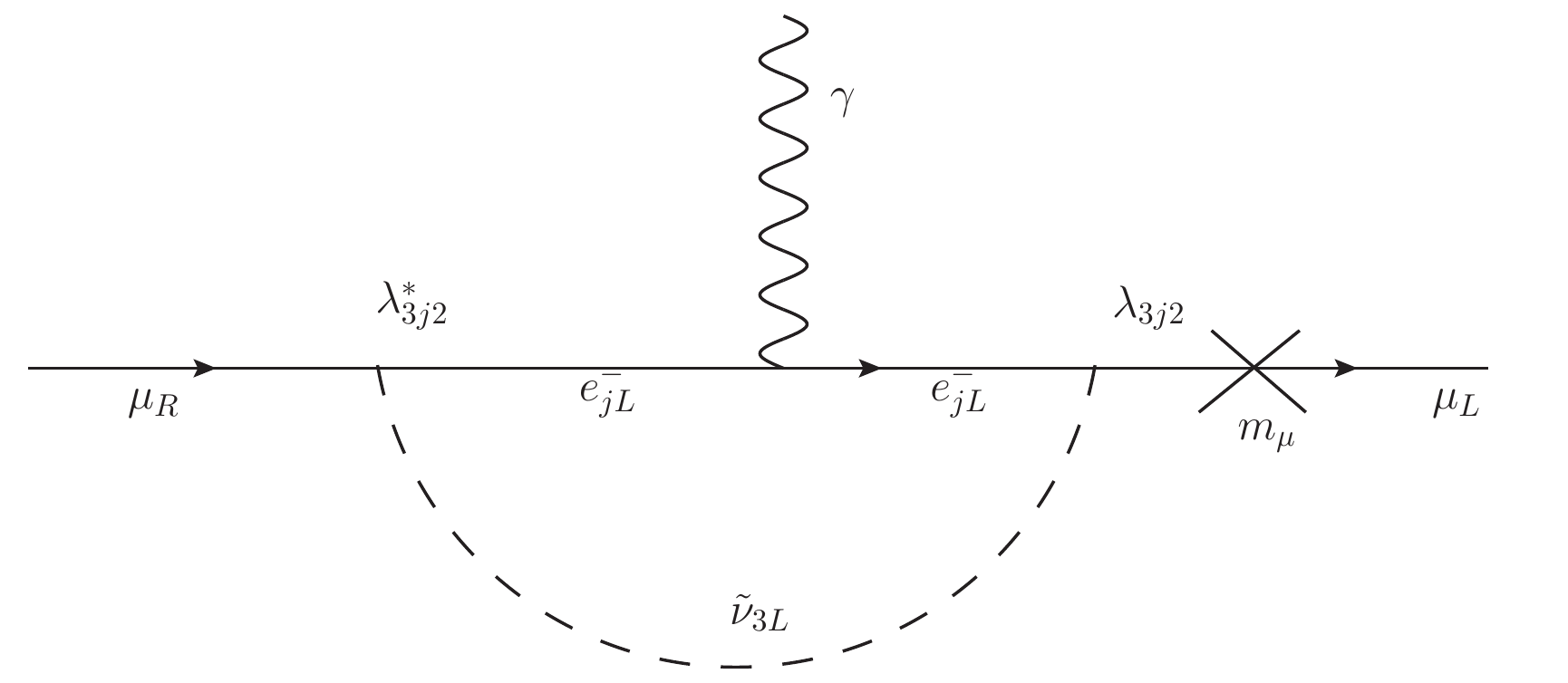} }}%
    \qquad
    {{\includegraphics[width=6cm]
    {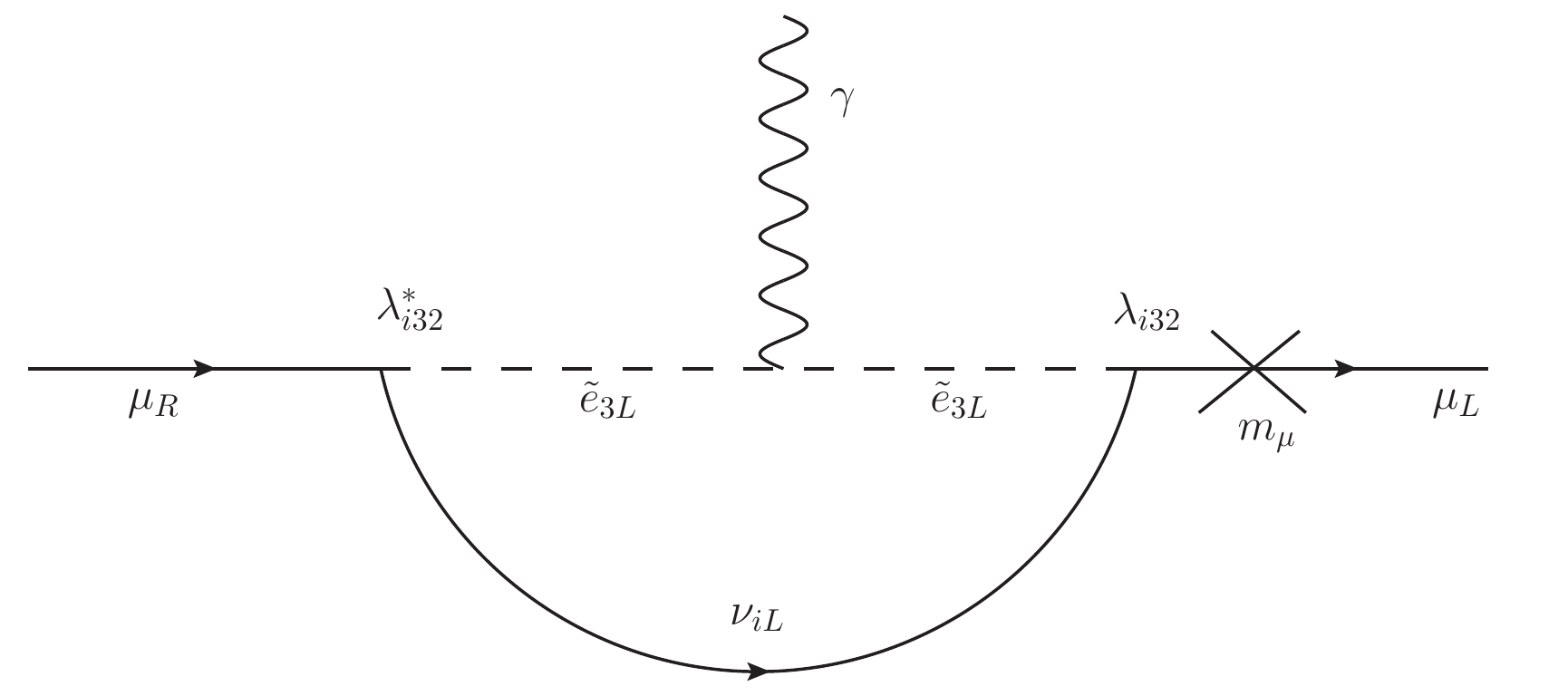} }}%
    \caption{The most dominant diagrams which contribute to 
    $(g-2)_{\mu}$ in RPV MSSM scenario.
    }
    \label{g-2}%
\end{figure}

Furthermore, as we have already discussed, in the absence of $R$-parity, 
the superpotential contains additional terms which are
lepton and baryon number violating. In the context of our analysis we will consider
only the following terms in the 
superpotential\footnote{The bounds on $\lambda^\prime$ operators are more stringent compared to the
bounds on $\lambda$ from neutrino masses. In addition, the presence of 
$\lambda^\prime$ operators will increase the direct production cross-section of
the sneutrinos subjecting to stronger constraints on the sneutrino
masses. Furthermore, we also assume baryon number is conserved. 
As a result, we confine ourselves to the $\lambda$-type couplings only.}
\begin{eqnarray}
W_{\cancel{R}_p}=W_{\text{MSSM}}+\frac{1}{2}\lambda_{ijk}\widehat L_i \widehat L_j 
\widehat E_k^c,
\end{eqnarray}
where $W_{\text{MSSM}}$ contains the usual MSSM superfields and $\widehat L$, 
$\widehat E^c$ are the left-chiral lepton and left-chiral anti-lepton superfields
respectively. Gauge invariance enforces $\lambda_{ijk}$ to be antisymmetric 
with respect to their first two indices. As a result, $\lambda_{ijk}=-\lambda_{jik}$.
These RPV terms in the superpotential yield the following terms in the Lagrangian
in the form as
\begin{eqnarray}
\mathcal L &=& -\frac{1}{2}\lambda_{ijk}\Big[\widetilde\nu_{iL}\bar l_{kR}l_{jL}+\widetilde l_{jL}
\bar l_{kR}\nu_{iL}+\widetilde l_{kR}^{*}\bar\nu_{iR}^c l_{jL}-(i\leftrightarrow j)
\Big] \nonumber \\
&+& h.c. 
\label{eq:l1}
\end{eqnarray}
In the four component notation, the terms in Eq.~\ref{eq:l1} 
which contribute to
$(g-2)_{\mu}$ can be explicitly written as 
\begin{eqnarray}
\mathcal L &\subset& -\lambda_{ij2}\Big[\widetilde \nu_{iL}\bar\mu P_L l_j+\widetilde l_{jL}
\bar\mu P_L \nu_i\Big]\nonumber \\
&-& \lambda_{i2k}\Big[\widetilde \nu_{iL}\bar l_K P_L\mu+\widetilde l_{kR}^{*}
\bar \nu_i^c P_L \mu\Big]+h.c.
\label{4comp}
\end{eqnarray}
In our scenario, we assume the first two generations of right chiral sleptons to
be heavy to evade constraints appearing from neutrino masses and low energy
observables in the presence of order one $\lambda$'s as elaborated later. However,
the left chiral charged sleptons/sneutrinos can be light and still avoiding bounds
from direct collider constraints. Therefore, in addition to the $\Delta a_{\mu}$
contribution coming from RPV operators, we also have sizeable contribution from the
RPC section. The relevant diagrams contributing to $(g-2)_{\mu}$ are shown in 
Fig.~\ref{g-2}.
The generic expression of $(g-2)_{\mu}$ 
in the context of RPV MSSM are written as \cite{Kim:2001se}
\begin{eqnarray}
\Big[a_{\mu}^{\lambda}\Big]_{\text{RPV}}&=& \frac{m_{\mu}^2}{96\pi^2}
\Bigg[|\lambda_{23k}|^2\frac{2}{m^2_{\widetilde\nu_{\tau}}}+|\lambda_{3k2}|^2
\Bigg\{\frac{2}{m^2_{\widetilde\nu_{\tau}}}-\frac{1}{m^2_{\widetilde\tau_L}}
\Bigg\}\nonumber \\
&-&|\lambda_{k23}|^2 \frac{1}{m^2_{\widetilde\tau_R}}
\Bigg],
\label{amu}
\end{eqnarray}
where $m_{\widetilde\tau_L}$ and $m_{\widetilde\tau_R}$ are the left and right 
chiral stau masses respectively while $m_{\widetilde\nu_{\tau}}$ is the 
tau-sneutrino mass. In the limit where all the relevant third generation slepton masses are
considered to be equal, i.e., $m_{\widetilde\tau_L}=m_{\widetilde\tau_R}
=m_{\widetilde\nu_{\tau}}
=\widetilde m$, then Eq.~(\ref{amu}) reduces to the following simplified form
\begin{eqnarray}
\Big[a_{\mu}^{\lambda}\Big]_{\text{RPV}} &=& \frac{m_{\mu}^2}{32\pi^2\widetilde m^2}\Bigg[
\frac{1}{3}|\lambda_{312}|^2+\frac{2}{3}|\lambda_{321}|^2+\frac{1}{3}|\lambda_{323}|^2\nonumber \\
&+&|\lambda_{322}|^2 -\frac{1}{3}|\lambda_{123}|^2\Bigg].
\label{amu-simp}
\end{eqnarray}
An important observation is except $\lambda_{322}$, all the other RPV couplings
come with a factor less than one. Our goal is now to study the present bounds
on these couplings and to make sure if such an order one $\lambda$ can be considered.
{We note in passing that in the present work we have taken into account all
the contributions to anomalous magnetic moment of the muon coming from both
the RPC as well as RPV MSSM.}
\section{Bounds on RPV couplings}
\label{bounds}
In the paradigm of SM, the lepton flavor violating (LFV) processes occur at a 
negligible rate due to the smallness of the neutrino masses. As a result, they are
sensitive probe of new physics and can be used to place bounds on $\cancel{R}_p$
couplings\footnote{For bounds on trilinear R-parity violating couplings see 
Refs.~\cite{Dreiner:2012mx,Dreiner:2006gu,Dreiner:2010ye,Saha:2002kt,
Saha:2003tq,Kundu:2004cv,deGouvea:2000cf,Chaichian:1996wr,Bhattacharyya:1995pr,
Agashe:1995qm,Ghosh:1996bm,Ghosh:2001mr,Bhattacharyya:1998bx,Abada:2000xr,
Allanach:2003eb,Littenberg:2000fg}.}. In order to disentangle the effects of $\cancel{R}_p$ interactions
from the effects emerging from the possible flavor non-universalities in the
scalar lepton sector, we assume that the slepton mass matrices are diagonal with
first two generations having equal masses. The possible sources of LFV are noted
down in the following processes. 
\begin{itemize}
\item {Lepton flavour violating radiative decays of charged leptons} : The 
$\cancel{R}_p$ interactions can in principle generate LFV decays of charged leptons, 
such as $l_i\rightarrow l_j\gamma$ through one loop diagrams \cite{Adam:2011ch}.

\item {Lepton flavour violating decays of $\mu$ and $\tau$ into three charged 
leptons} : The LFV decays like $l_m^- \rightarrow l_i^- l_j^- l_k^+$, where 
$l_m=\mu, \tau$ can be mediated at the tree level through $t$ and $u$-channel 
sneutrino exchanges when the involved leptons posses non-zero $\lambda$ type $\cancel{R}_p$ couplings.
The non-observation of these processes results in bounds on $\lambda_{nmi}
\lambda_{njk}^*$, where the sneutrino carries the index $n$ \cite{Agashe:2014kda}.

\item {Muon to electron conversion in nuclei} : $\mu^- \rightarrow e^-$ conversion
in a nucleus is normally induced by $\lambda\lambda^{\prime}$ or $\lambda^{\prime}
\lambda^{\prime}$ couplings\footnote{ For a theoretical 
calculation of this process in R-parity-violating SUSY models, we 
refer Ref.~\cite{Lee:1984kr}.}. However, $\mu^-\rightarrow e^-$ conversion in a nucleus
can also proceed through photon penguin diagrams. The associated bounds can be
much stronger than the ones extracted from the previously mentioned processes.
The non-observation of these processes can be translated into bounds on
$\lambda\lambda$ couplings \cite{Dohmen:1993mp,Honecker:1996zf,Bertl:2006up}.

\item {Charged current universality and bounds from $R_{\tau}/R_{\tau\mu}$} :
One should also take into account bounds from charged current universality which
results in single bounds on the $\lambda$ couplings. Similar bounds can also be
obtained from the ratio $R_{\tau}=\Gamma(\tau\rightarrow e\nu\bar\nu)/\Gamma(\tau
\rightarrow \mu\nu\bar\nu)$ and $R_{\tau\mu}=\Gamma(\tau\rightarrow\mu\nu\bar\nu)/
\Gamma(\mu\rightarrow e\nu\bar\nu)$ \cite{Adam:2011ch}.
\end{itemize}

We now tabulate the bounds on the relevant $\cancel{R}_p$ couplings from the 
non-observation of the processes as mentioned earlier.
\begin{table}[h!]
\centering
\begin{tabular}{||c| c| c| c| c||} 
 \hline
 $\cancel{R}_p$ couplings & $l_i\rightarrow l_j\gamma$ 
 & $l_i\rightarrow 3 l_j$ & $\tau\rightarrow l_i P/\mu-e$ 
 & $l_i\rightarrow l_j l_k l_k$\\ [0.5ex] 
 \hline\hline
 $|\lambda^*_{312} \lambda_{322}|$ & $2.3\times 10^{-3}$ & $8.2\times 10^{-4}$ 
 & $1.6\times 10^{-4}$ & -\\ 
 $|\lambda^*_{321} \lambda_{322}|$ & $3.8\times 10^{-4}$ & $4.1\times 10^{-4}$ 
 & $1.2\times 10^{-4}$ & -\\
 $|\lambda^*_{323} \lambda_{322}|$ & - & - & - & $2.4\times 10^{-3}$ \\ [1ex] 
 \hline
\end{tabular}
\caption{Bounds on $\cancel{R}_p$ couplings from low energy experiments 
\cite{Adam:2011ch,Agashe:2014kda,Dohmen:1993mp,Honecker:1996zf,Bertl:2006up} 
with specific benchmark point as shown in Ref.\cite{Dreiner:2012mx}.}
\label{table:1}
\end{table}
All these limits are obtained from BP1 of Ref~\cite{Dreiner:2012mx}, where the first two 
generations are considered to be heavy with masses around 1 TeV and the third generation 
is light. Making the first two generations masses heavier would further relax the 
bounds on $\cancel{R}_p$ couplings. However, we take a more conservative 
approach here and use the strongest limits. In addition, from the 
charge current universality one finds $|\lambda_{123}|
\sim 0.049\times {m_{\widetilde\tau_R}}/{100\text{GeV}}$. Since, in our
framework, the third generation is considered to be light hence the bounds
on the particular $\cancel{R}_p$ operators turn out to be stringent and should
be respected. From $R_{\tau}$ and $R_{\tau\mu}$ one finds $|\lambda_{322}|<0.07\times
m_{\widetilde\mu_R}/100~\text{GeV}$. Hence, in our scenario this bound can be
readily relaxed by assuming large mass for the second generation sleptons which we
have considered. As mentioned earlier, this bound also translates to a lower
bound on $\widetilde e_R$ as they are considered to be degenerate with $\widetilde\mu_R$. 

From the above discussion and Table~\ref{table:1}, it is 
conspicuous that only one of the $\cancel{R}_p$ violating
operator can be large ($\mathcal O(1)$) satisfying the above mentioned
constraints. We choose it to be $\lambda_{322}$.
\begin{itemize}
\item {{Bound on $\cancel{R_p}$ couplings from neutrino mass}} : Neutrino masses provide 
serious constraints on trilinear $\cancel{R}_p$ couplings. In this section we will compute the
impact of neutrino masses on $\lambda_{322}$. These couplings generate neutrino masses
radiatively (see Fig.~\ref{trilinear-numass}) and the generic expression 
is noted down as \cite{Grossman:1998py,Grossman:2003gq}
\begin{figure}%
    \centering
    {{\includegraphics[width=6cm]
    {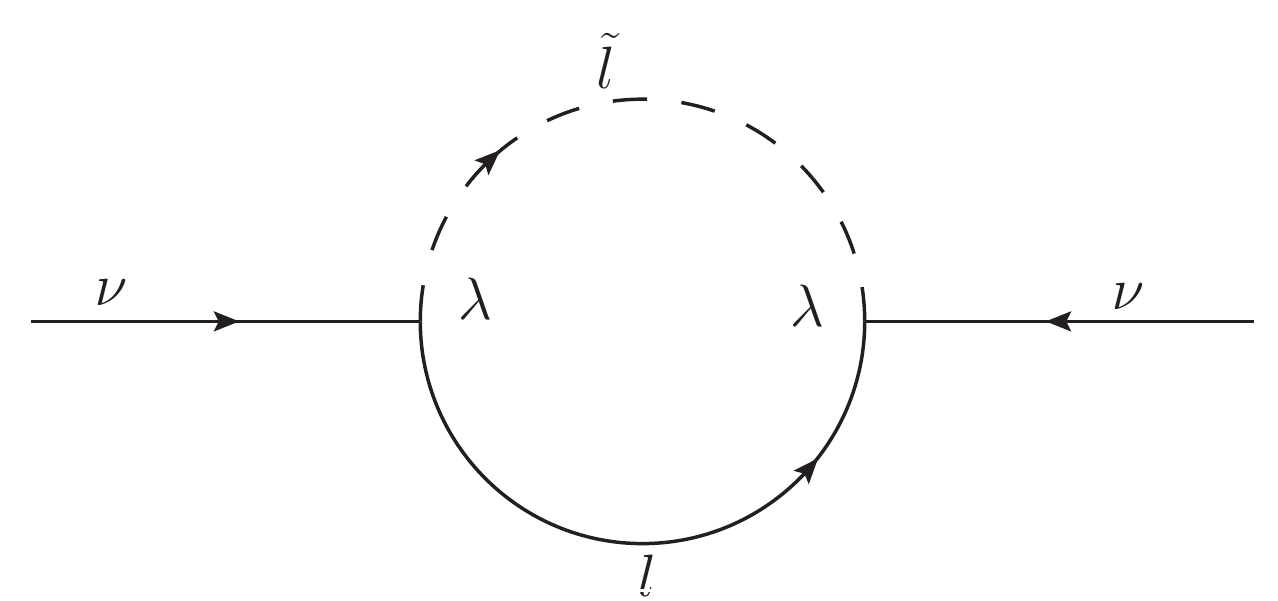} }}%
    \caption{Trilinear $\cancel{R}_p$ violating contribution to neutrino
    masses.
    }
    \label{trilinear-numass}%
\end{figure}
\begin{eqnarray}
(m_{\nu})_{qm} &\simeq& \frac{1}{32\pi^2}\sum_{l,p} \lambda_{qlp}\lambda_{mpl}m_l
\sin 2\phi_l \ln\Big(\frac{M^2_{p1}}{M^2_{p2}}\Big),
\label{numass}
\end{eqnarray}  
where $m_l$ is the mass of the lepton, $\phi_l$ is the mixing angle obtained by
diagonalising the charged slepton mass squared matrix, which takes the form
\begin{eqnarray}
\sin 2\phi_l &=& \frac{2 A m_l}{\sqrt{(L^2-R^2)^2+4 A^2 m_l^2}}, 
\end{eqnarray}
where $L^2\equiv (m^2_{\widetilde L})_{ll}+(T_3-e\sin^2\theta_W)m_Z^2\cos 2\beta$,
$R^2\equiv (m^2_{\widetilde E})_{ll}+(e\sin^2\theta_W)m_Z^2\cos 2\beta$, with
$T_3=-1/2$ and $e=-1$ for the down-type charged sleptons, and the effective
trilinear scalar coupling term is denoted as $A\equiv (A_E)_{0ll}-\mu\tan\beta$.
$M_{p1}$ and $M_{p2}$ are slepton mass eigenstates obtained by diagonalising
the slepton mass squared matrix. The trilinear $\cancel{R}_p$ operator under 
consideration, i.e., $\lambda_{322}$ gives mass to the (33) element of the 
neutrino mass matrix. As a result, Eq.~(\ref{numass}) can be simplified to
\begin{eqnarray}
(m_{\nu})_{33}&\simeq& \frac{1}{16\pi^2}|\lambda_{322}|^2 \frac{A m_{\mu}^2}
{\sqrt{(L^2-R^2)^2+4 A^2 m_l^2}}\nonumber \\
&&\ln\Big(\frac{M^2_{p1}}{M^2_{p2}}\Big).
\label{mass33}
\end{eqnarray}
Considering the central values for the neutrino mass squared and mixing parameters 
\cite{Forero:2014bxa} (with the choice, CP violating phase $\delta=0$), the central value of the 33
element of the neutrino mass matrix for normal and inverted hierarchy becomes
\begin{eqnarray}
(m_{\nu})_{33}^{NH}&=&0.023~\text{eV}, \nonumber \\
(m_{\nu})_{33}^{IH}&=&0.031~\text{eV}
\label{niv} 
\end{eqnarray}
From Eq.~(\ref{mass33}), it is straight-forward to show that in the limit
$(R^2\gg L^2)\equiv \widetilde m^2\gg A^2$, the same equation gives the following
bound on the $A$ parameter as
\begin{eqnarray}
A \ln\Bigg(\frac{M_{p1}}{M_{p2}}\Bigg) \leq \mathcal O(10)~\text{GeV},
\label{smallA}
\end{eqnarray}
for $\mathcal O(1)$ $\lambda_{322}$ and $\widetilde m$, i.e., the first 
two generations of right slepton masses are in the ballpark of 
$\mathcal O(10~\text{TeV})$. Therefore, we
observe that in order to consider $\lambda_{322}\sim \mathcal O(1)$, one needs
to satisfy Eq.~(\ref{smallA})\footnote{The issues pertaining to neutrino masses and 
muon $(g-2)$ anomaly in the framework of RPV SUSY can also be found in 
Refs.~\cite{Adhikari:2001ra,Adhikari:2001sf}.} which invokes a cancellation 
between the soft SUSY breaking $A$ term in the charged slepton sector and the 
$\mu$ term. 
\end{itemize}
\section{Numerical Analysis and benchmarks}
From the previous discussions, it is clear that only $\lambda_{322}$ plays
dominating role in ameliorating the tension between the observed muon anomalous
magnetic moment and the SM expectation. In the limit when only $\lambda_{322}$
is non-zero, whereas all the other trilinear $\cancel{R}_p$ violating couplings 
are vanishingly small, Eq.~(\ref{amu-simp}) further simplifies to \cite{Kim:2001se}
\begin{eqnarray}
\Big[a_{\mu}^{\lambda}\Big]_{\text{RPV}}&\simeq& 34.9\times 
10^{-10}\Bigg(\frac{100~\text{GeV}}{\widetilde m}\Bigg)^2
|\lambda_{322}|^2.
\end{eqnarray} 
It is conspicuous that an order one coupling can explain the muon anomalous magnetic
moment event within 1$\sigma$ of the central measured value. 

In order to have a complete and concrete picture, we use
the trilinear R-parity violating model implemented in {\tt SARAH v-4.4.6} 
\cite{Staub:2008uz,Staub:2015kfa}. The spectrum
has been generated using {\tt SPheno v-3.3.3} \cite{Porod:2003um,Porod:2011nf}. 
{\tt FlavorKit} \cite{Porod:2014xia} is used to ensure that the benchmark points are consistent with 
all relevant flavour violating observations. We fix the following parameters, such
as, the bino mass parameter $M_1=300~\text{GeV}$, the wino mass parameter $M_2=1.7~\text{TeV}$,
the Higgsino mass term $\mu=200~\text{GeV}$, the gluino mass $M_3=1.5~\text{TeV}$,
$\tan\beta=v_u/v_d=20$ and the mass of the CP-odd Higgs $M_A=400~\text{GeV}$.
$\lambda_{322}$ is varied from 0.5 to 1.2 keeping all other $\cancel{R}_p$ couplings 
to zero\footnote{In order satisfy the muon $(g-2)$ and the LEP bound on the 
sneutrino mass simultaneously leads to $\lambda_{322} \ge$ 0.5, however we 
restrict ourselves within $\lambda_{322} \le$ 1.2.}. We also vary the soft mass squared term of the slepton doublet
in the limit $3\times 10^{4}~\text{GeV$^2$}\leq (m_{33}^L)^2\leq 2.5\times 10^{5}~
\text{GeV$^2$}$ and chalk out the parameter space by putting the $\Delta a_{\mu}$ 
constraints within $1 ~\& ~2\sigma$ regime.
\begin{figure}%
    \centering
    {{\includegraphics[width=8cm]
    {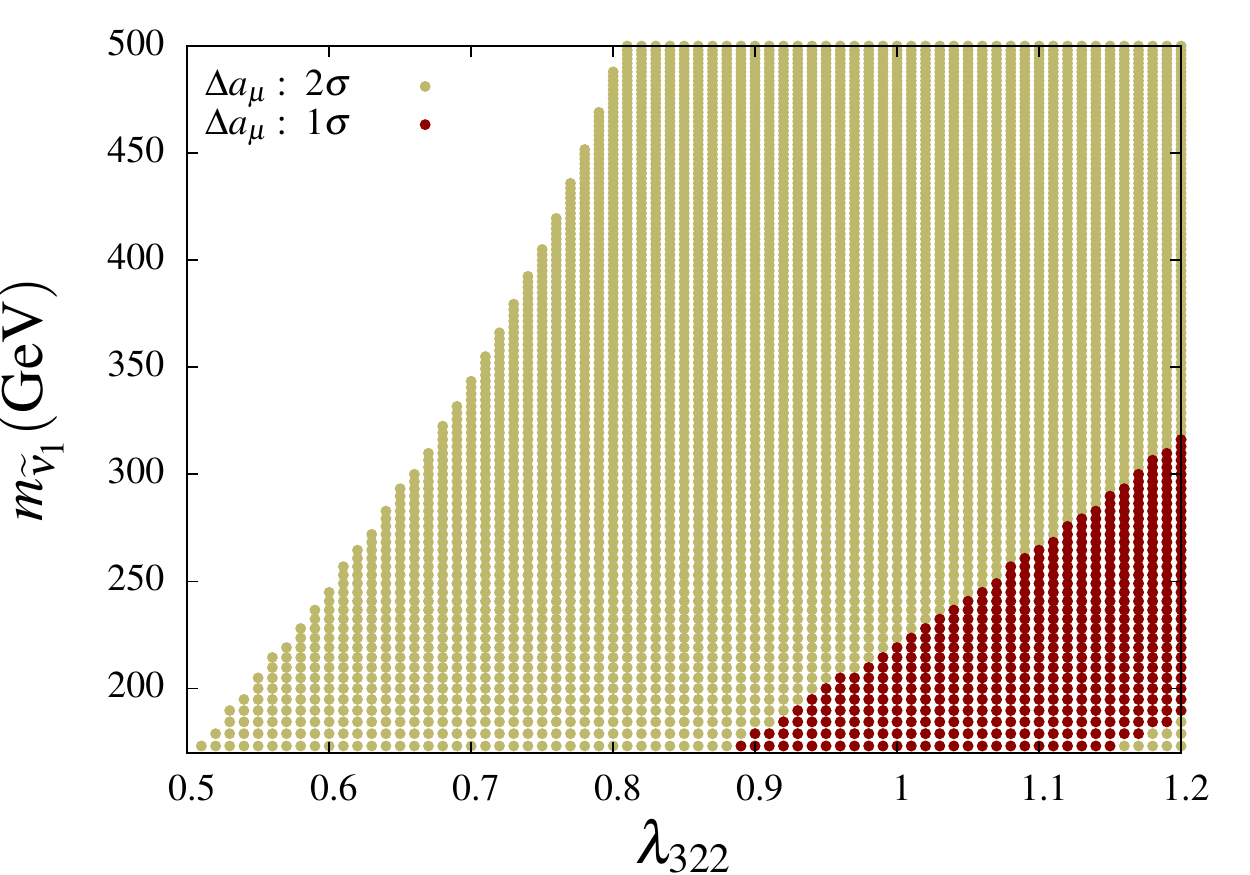} }}%
    \caption{1 and 2$\sigma$ limits on $\Delta a_{\mu}$ are shown in
red and yellow colours respectively in the $m_{\widetilde\nu_1}-\lambda_{322}$ 
plane where $m_{\widetilde\nu_1} \equiv m_{\widetilde\nu_\tau}$. 
}
    \label{muong}%
\end{figure}

In Fig.~\ref{muong} we show the 1 and 2$\sigma$ constraints on $\Delta a_{\mu}$
in the $m_{\widetilde\nu_1}-\lambda_{322}$ plane where ${\widetilde\nu_1}$ is the 
lightest mass eigenstate of the sneutrinos with $\widetilde\nu_1\equiv\widetilde\nu_\tau$. 
We observe that the 1$\sigma$ regime
of this parameter can be reached for large values of $\lambda_{322}\geq 0.9$ and
in the low mass limit of $m_{33}^L$, which also controls the left sneutrino and
left charged slepton masses. From Eq.~\ref{4comp}, a nonzero $\lambda_{322}$ implies 
the tau-sneutrino decaying to a $\mu^+\mu^-$ final state.
It is important to note that pair produced tau-sneutrinos decaying 
via 4$\mu$ final state has not been 
looked at by both the ATLAS and CMS collaborations. However, experiments have 
looked for pair production of $\widetilde\mu_{L}$ which decays to the 
$\mu$ and $\chi^0_1$, and placed a mass limit on $\widetilde\mu$. Now, in our 
case, the stau $\widetilde \tau_{L}$ associated to $\widetilde\nu_\tau$ can 
decay to $\mu\nu_{\mu}$, thus a pair produced stau would give same 
final state topology through the same RPV operator. Hence, the present 
bound on $\widetilde\mu$ can be attributed to $\widetilde\tau$ (and hence 
$\widetilde\nu_\tau$) in our case. The present lower bound on 
$\widetilde\mu$ stands at 300 GeV \cite{Aad:2014vma,Khachatryan:2014qwa}, and thus 
this bound can also be mapped to an lower bound on $\widetilde\nu_{\tau}$ mass.
However, it is also important to note that by reducing the branching ratio of $\widetilde\nu_{\tau}
\rightarrow\mu^+{\mu^-}$, one can relax the bound considerably. For example, we check 
that for ${\rm BR}(\widetilde\nu_{\tau}\rightarrow\mu^+{\mu^-}) \sim$ 70\%, the 
bound on the sneutrino mass reduces to 250 GeV, while for 
${\rm BR}(\widetilde\nu_{\tau}\rightarrow\mu^+{\mu^-}) \sim$ 50\% the bound 
on the same is around 220 GeV.

In our scenario, the partial decay width of the sneutrino 
(in this case the tau sneutrino)
decaying to $\ell^+\ell^-$ is given by 
\begin{eqnarray}
\Gamma(\widetilde\nu_i\rightarrow l_j^+ l_k^-)&\simeq& \frac{1}{16\pi}\lambda_{ijk}^2
m_{\widetilde\nu_i}.
\end{eqnarray} 
Further, if kinematically allowed, the sneutrino 
can also undergo a two body decay with a tau neutrino and a neutralino or a tau 
lepton associated with a chargino in the final state. The neutralino and chargino
would also undergo a three body decay in the RPV framework. The two body decay 
widths of the sneutrino are noted below \cite{Grossman:1997is}
\begin{eqnarray}
\Gamma (\widetilde\nu\rightarrow\widetilde\chi_j^0\nu)&=&\frac{g^2 |Z_{iZ}|^2 
m_{\widetilde\nu}}{32\pi\cos^2\theta_W}B(m^2_{\widetilde\chi_j^0}
/m^2_{\widetilde\nu}),\nonumber \\
\Gamma (\widetilde\nu\rightarrow\widetilde\chi^+ \ell^-)&=&\frac{g^2 |V_{11}|^2
m_{\widetilde\nu}}{16\pi}B(m^2_{\widetilde\chi^+}/m^2_{\widetilde\nu}),
\end{eqnarray}
where $V_{11}$ is one of the mixing matrix elements in the chargino sector and 
$Z_{jZ}$ is the neutralino mixing matrix element. The $B$ function is defined as
$B(x)=(1-x)^2$. In the presence of large $\lambda_{322}$, which is also motivated
from the perspective of fitting $\Delta a_{\mu}$, the partial decay width 
of the sneutrino decaying to a pair of leptons will dominate over the other 
decay modes.   
\begin{figure}%
   \centering
    {{\includegraphics[width=8cm]
    {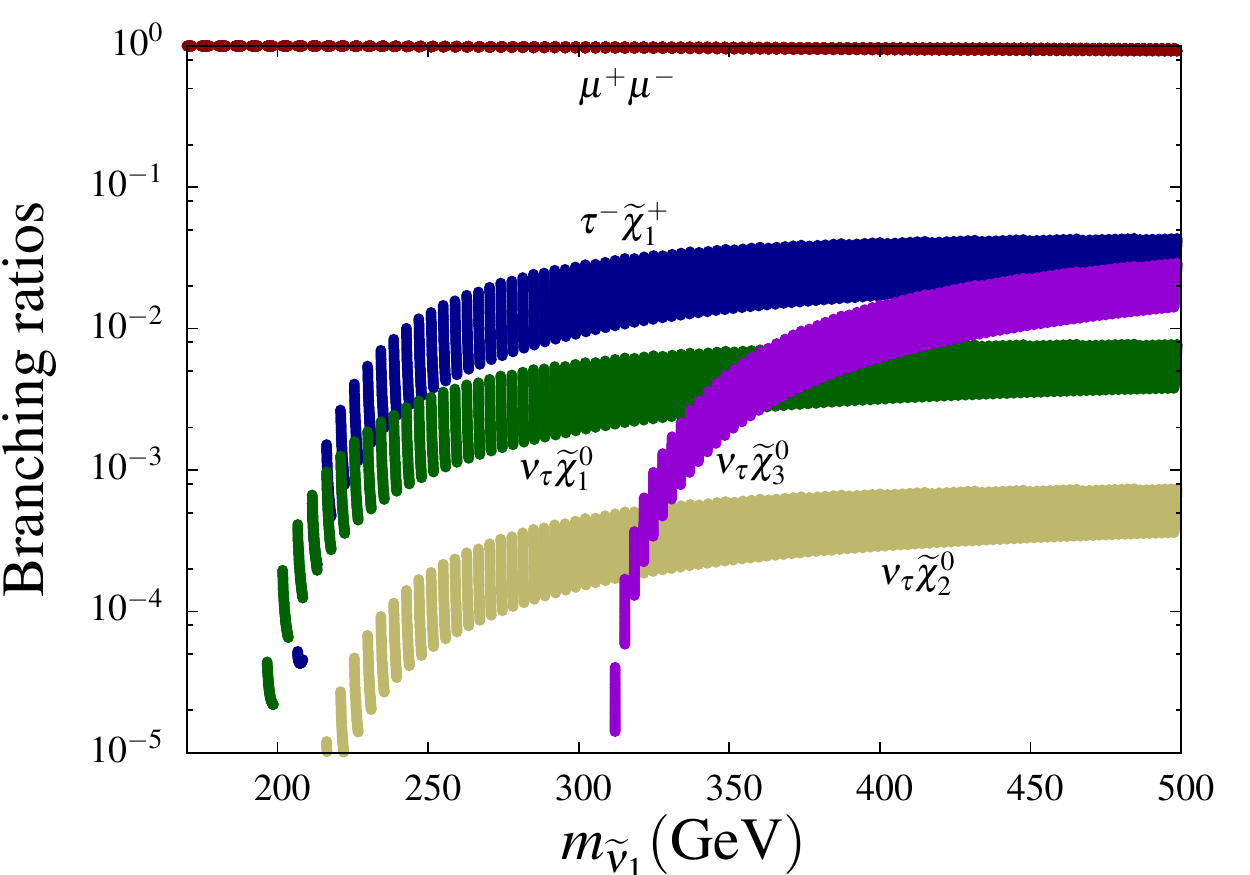} }}%
    \qquad
    {{\includegraphics[width=8cm]
   {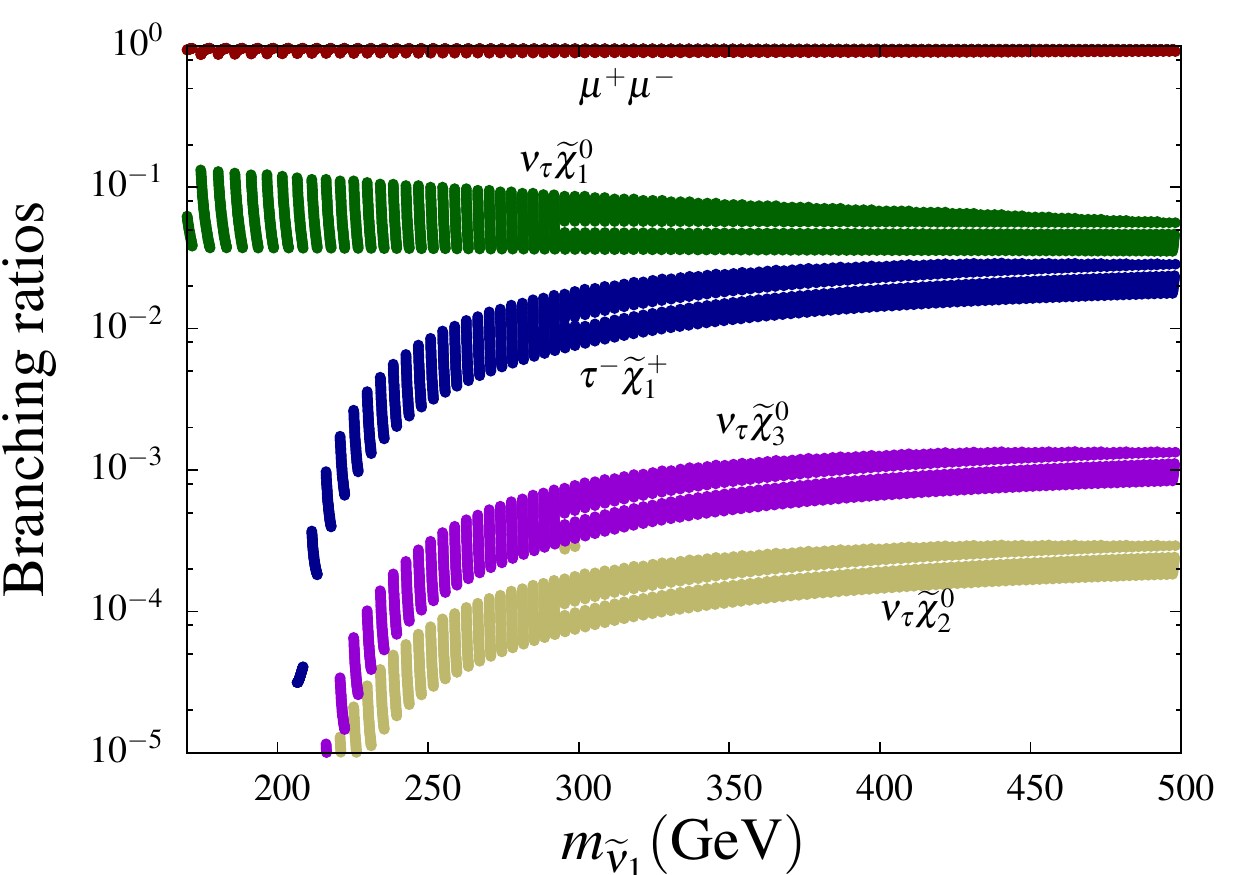} }}%
    \caption{Decay branching ratios of the lightest sneutrino for
    bino-line neutralino mass parameter of $M_1=300$ GeV and
    $M_1=10$ GeV respectively. The points are consistent within 
 2$\sigma$ error of the $\Delta a_{\mu}$ parameter.
    }
    \label{branching}%
\end{figure}
In Fig.~\ref{branching}, we portray the branching ratios of the lightest
sneutrino as a function of its mass. During this scan, we ensure that all 
the points satisfy the Higgs mass and branching ratio constraints and also the 
low energy experimental constraints. In addition, care has been taken in removing
all the tachyonic states from the scan. The points are also consistent within $2\sigma$ error
of the $\Delta a_{\mu}$ parameter. All the parameters are fixed at the previously
mentioned values expect for $M_1$. In the first column of Fig.~\ref{branching},
$M_1$ is fixed at 300 GeV, while in the lower panel $M_1$ = 10 GeV. 
As a result, sneutrino decays to charginos with associated leptons
and neutrino+neutralino final states are highly suppressed due to phase space consideration.
However, the bino-like neutralino mass parameter can be light (we choose it to be 
10 GeV). There are two major constraints
for light bino-like neutralino, for example, one has to check if Higgs partial decay width
into this channel is satisfied or not and secondly, in the RPV scenario, the light neutralino
can decay into final states involving fermions and can avoid constraints from its 
overproduction in the early universe \cite{Dreiner:2009ic,Chakraborty:2013gea}. 
In addition, the added advantage of this scenario
is the presence of light neutralino opens up new decay modes of the 
sneutrino. Thus the effective branching ratio of this sneutrino
decaying to two muon final state can be reduced. As a result, the branching ratio of the
stau decaying to $\mu\nu_\mu$ also reduces and thus relaxes 
the bound on the left handed stau mass. This in turn also implies 
the left-chiral sneutrino mass bounds can be relaxed further as 
both the masses are controlled by the same parameter. 
Before we proceed any further, 
let us give a brief outline of the search for heavy di-muon resonances 
at the Tevatron and LHC experiments. 

\begin{figure}%
    \centering
    {{\includegraphics[width=8cm]
    {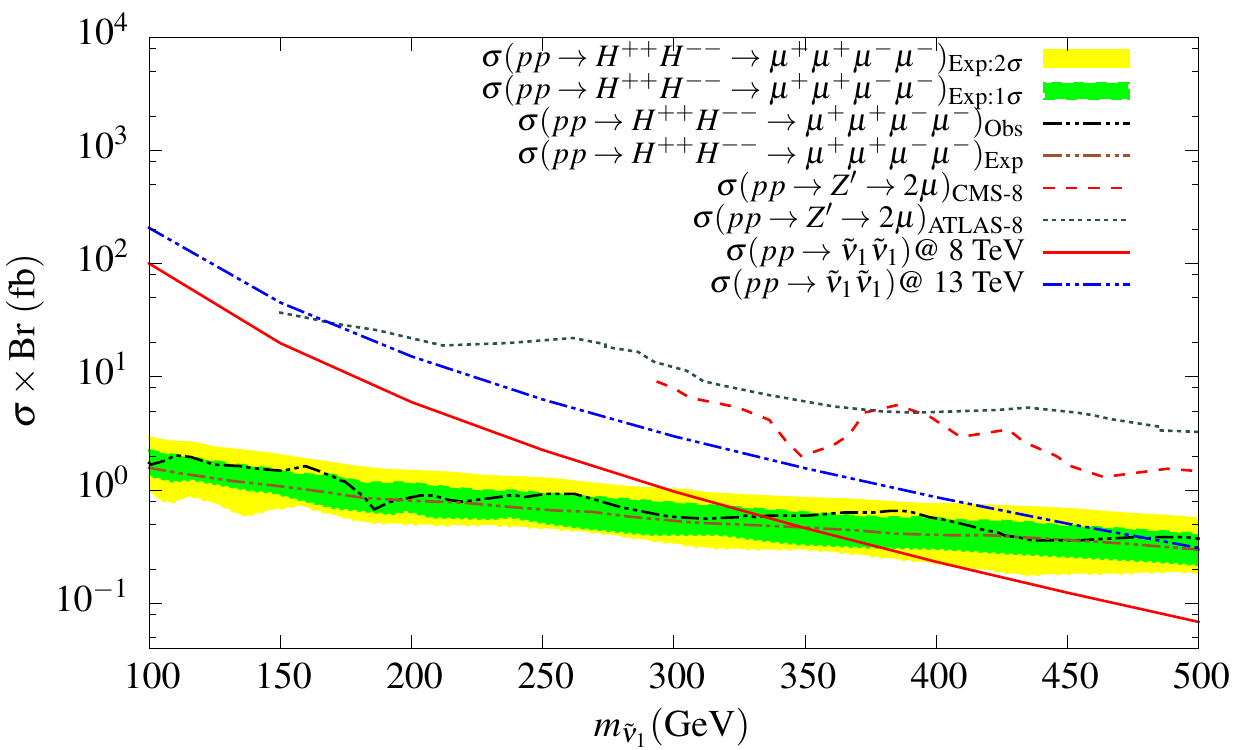} }}%
    \caption{Present 95\% C.L. upper limits on the $\sigma \times BR$ for 
different values of heavy resonance masses.  }
    \label{fig:sigmaBR}%
\end{figure}

Heavy resonances decaying to a pair of muons naturally comes in many 
extensions of the SM with additional gauge groups. Both the 
ATLAS and CMS collaborations have searched for the 
heavy spin-1 resonance $Z^\prime$ via di-muon final states at the 
7 and 8 TeV run of LHC \cite{Aad:2014cka,Khachatryan:2014fba,ATLAS:2012hmt}. Non-observation 
of signatures of the signal events 
leads to the 95\% C.L. upper limits on the production cross-section times 
branching ratios over a range of di-muon invariant masses. In Fig.~\ref{fig:sigmaBR}, 
the black dotted and red dashed lines indicate the corresponding limits 
obtained from the ATLAS and CMS collaborations by the LHC-8 data respectively. 
Moreover, the CDF collaboration at the Tevatron experiment has also performed a study of 
di-muon resonances from the direct production of a sneutrino or $Z^\prime$ with 
1.96 TeV data \cite{Aaltonen:2008ah,Aaltonen:2011gp}. However, we 
do not consider the bounds coming from the 
sneutrino production since it involves the $\lambda^\prime$ LQD coupling in the 
production process which we set to be zero throughout 
our analysis. We find that 
our di-muon resonances, shown in blue dashed double-dotted (13 TeV) 
and red solid (8 TeV) lines, have smaller cross-sections compared to the ATLAS and 
CMS limits as elaborated later in the text.

Furthermore, we also consider the present bounds obtained 
from exotic searches at the LHC with final state topologies similar to 
ours i.e., with four muons among which two are positively 
charged and two negatively charged \cite{ATLAS:2014kca}. The ATLAS collaboration has searched 
for doubly charged Higgs bosons decaying to a pair of same sign muons, thus 
giving rising to the same final state signature. We again translate the 
95\% C.L. upper limit on the production cross-section times branching ratio 
for heavy resonance mass from 100 to 500 GeV. The black dashed double-dotted 
line indicates 
the 95\% C.L. upper limit on the cross-section times branching ratio, while 
the green and yellow regions denote the 1$\sigma$ and 2$\sigma$ bands 
around the expected line shown in brown dashed double-dotted line. 
From Fig.~\ref{fig:sigmaBR}, we see that if 
one allows 2$\sigma$ fluctuations then sneutrino mass less than 290 GeV 
is excluded from this exotic search.

It is to be noted that, since we set $\lambda^\prime$ to zero, 
the sneutrinos are produced at the LHC via only Higgs boson and off-shell $Z$ 
mediation and cross-section naturally becomes much smaller compared to present 
upper bounds except the doubly charged Higgs boson search process. The bound 
on the di-muon mass and hence on the sneutrino mass stands roughly at 290 GeV, 
similar to what we obtain translating the LHC bounds on the sleptons from the 
direct searches. Keeping all these bounds in mind, in 
Table~\ref{bptable2}, we show the benchmark 
points pertaining to two relevant scenarios
under consideration. The parameters which are fixed are $\tan\beta=20$, $\mu=200$ GeV,
$M_A=400$ GeV, $M_2=1.7$ TeV, $M_3=1.5$ TeV, $A_t=-1.9$ TeV, $\lambda_{322}=1.2$,
$(m^2_L)_{33}=8.92\times 10^4~\text{(GeV)}^2$ and $1.1\times 10^5~\text{(GeV)}^2$ 
respectively. $M_1$ is fixed at 10 GeV (BP1) and 300 GeV (BP2) respectively. 
Obtained $\Delta a_\mu$ is within 2$\sigma$ error bar of the central value.


\vspace*{0.3cm}
\begin{table}
\begin{center}
\begin{tabular}{|c|c|c|}
\hline
Point & BP1 & BP2 \\
\hline
\hline
\multicolumn{3}{|c|}{Mass spectrum} \\
\hline
\hline
$m_h$ (GeV)  & 124.2  & 124.3 \\
\hline
$M_{H^{0}}$ (GeV)  &  413   &  410   \\
\hline
$M_{H^{\pm}}$ (GeV)  &  421   &  418   \\
\hline
$m_{\widetilde g}$ (GeV)  &  1622  & 1622 \\
\hline
$m_{\widetilde t_1}$ (GeV) & 835 & 835  \\
\hline
$m_{\widetilde \nu_1}$ (GeV) &  291  &  323 \\
\hline
$\widetilde\chi_1^{\pm}$  (GeV) &  204   &  204  \\
\hline
$\widetilde\chi_2^{\pm}$  (GeV)  & 1711  & 1711  \\
\hline
$\widetilde\chi_1^0$ (GeV)   & 9  & 310 \\
\hline
$\widetilde\chi_2^0$ (GeV)  & 206  & 208  \\
\hline
$\widetilde\chi_3^0$ (GeV)  & 210  & 309  \\
\hline
$\widetilde\chi_4^0$ (GeV)  &  1711 & 1711 \\
\hline
$BR(b\to s\gamma)\times 10^{4}$ & 2.57  & 2.57 \\
\hline
$BR(B_s\to \mu^+\mu^-)\times 10^{9}$ & 3.96 &  3.98 \\
\hline
$\Delta a_{\mu} \times 10^{10} $  & 19.6  & 19.7 \\
\hline
\end{tabular}
\caption{Mass spectrum and a few observables for the two benchmark points.}
\label{bptable2}
\end{center}
\end{table}
\section{Collider Analysis}

Search for the new physics signatures with multiple
leptons has always been considered as the
golden channel mostly due to the cleanliness of the
final state topology. Both the ATLAS and CMS
collaborations at the LHC have searched for new
resonances via lepton-rich signatures in the
context of R-parity violating MSSM 
\cite{Aad:2015pfa,CMS:yut,Chatrchyan:2013xsw}.
However, the final state topologies that have been
studied by the CDF collaborations at the
Tevatron and ATLAS collaborations at the LHC
includes heavy neutral particles
decaying to $e\mu$, $e\tau$, or $\mu\tau$ \cite{Aaltonen:2010fv,Aad:2015pfa}.
From their analysis, we find that the $e\mu$ channel provides
the best sensitivity due to better resolution
for the electrons and muons (see Fig.2 of Ref.~\cite{Aad:2015pfa}). 
In this paper, we study possibly the cleanest 
final state topology which contains 
four isolated muons, which comes from the pair production of 
lightest sneutrino ($\widetilde\nu_1\equiv\widetilde\nu_\tau$) 
which subsequently decays to 
a pair of muons through a non-zero $\lambda_{322}$ R-parity violating 
coupling. We reiterate that this channel is also interesting from the 
perspective of ${(g-2)}_\mu$ anomaly.

\begin{figure}[h!]
    {{\includegraphics[height=7cm,width=8.5cm]
   {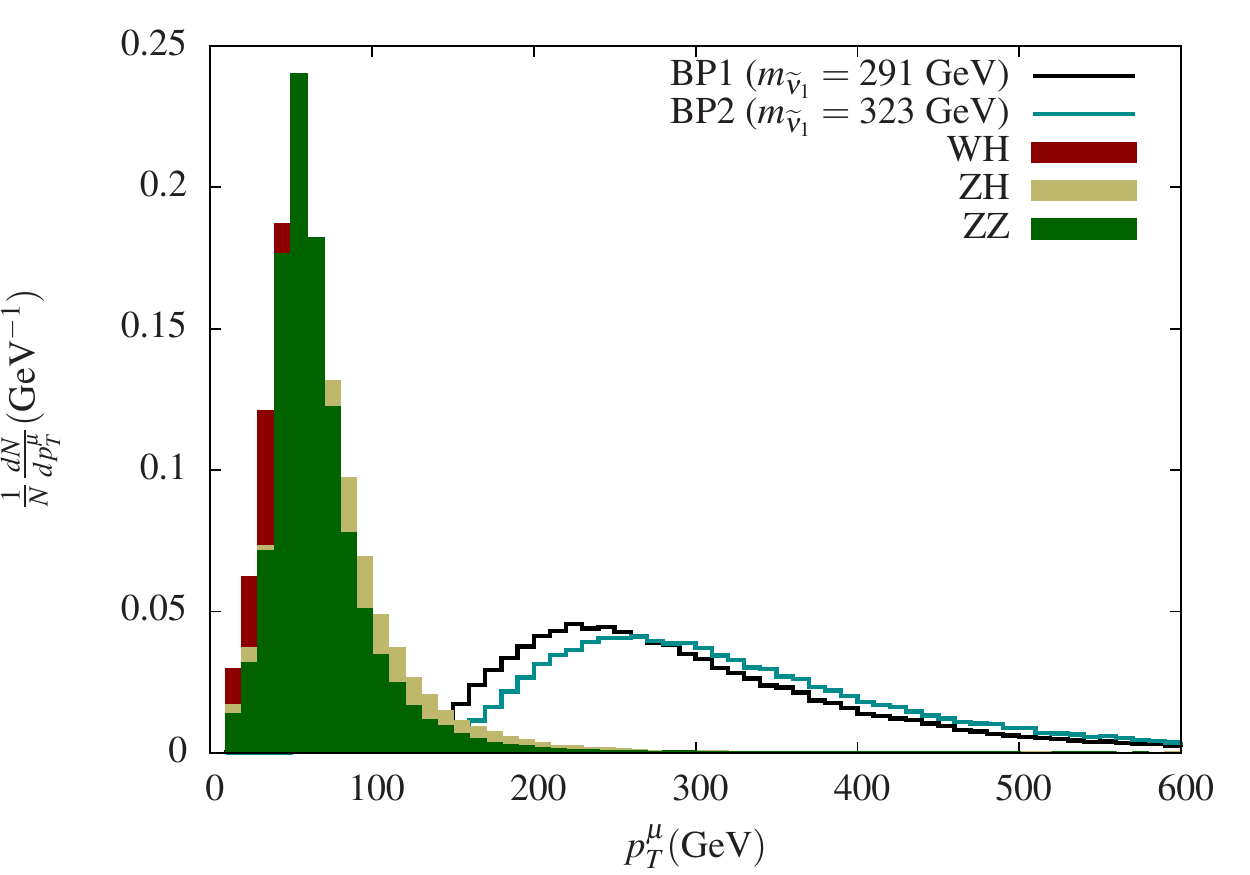} }}%
    \qquad
    {{\includegraphics[height=7cm,width=8.5cm]
    {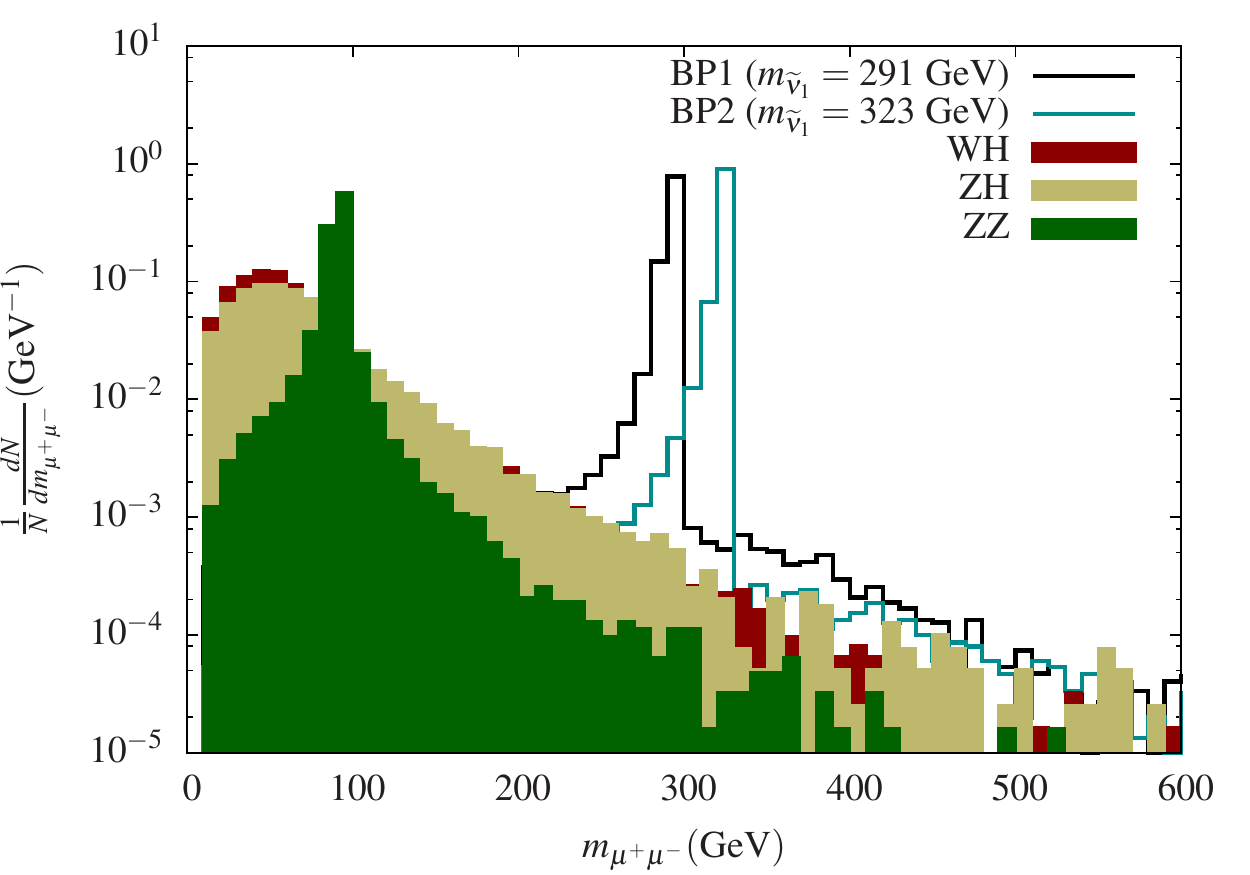} }}%
    \caption{ Upper: The $p_T$ distribution of the leading muon for the 
two benchmark points BP1 and BP2 along with the three 
dominant SM backgrounds. The signal muons are seen to be harder 
compared to that of the SM ones. Lower: Di-muon invariant mass 
distribution for the signal and background events. Sharp resonance 
peaks can be observed for the signal benchmark points, while a clear 
peak at $M_Z \sim $ 91 GeV is also visible for the mass distribution.}
    \label{minv}%
\end{figure}

We perform the collider analysis for the two benchmark points 
already introduced. We generate signal events 
using {\tt MadGraph (v5 2.2.2)} \cite{Alwall:2014hca} where the 
main sneutrino pair production channel involves $Z$ mediation. We 
then pass the events to {\tt PYTHIA (v 6.4.28)} \cite{pythia6} for hadronization and 
showering with {\tt CTEQ6L1} \cite{Pumplin:2002vw} 
parton density function. The final state of interest contains four isolated muons with 
no real source of missing energy. The possible SM backgrounds that 
can mimic the signal topology are as follows: $(i)$ SM 
Higgs boson production via 
gluon-gluon fusion, vector boson fusion, associated production 
processes with $H \to ZZ^{*} \to 4\mu$ final state. 
$(ii)$ Direct production of pair of SM gauge bosons i.e., $WW$, $WZ$ and 
$ZZ$ with $W/Z$ decay leptonically. 
$(iii)$ $Z+{\rm jets}$ and $t\bar t$ processes\footnote{
We check that processes like $t\bar tZ$, $t\bar tH$ with $H \to ZZ^{*}$, triple gauge 
boson productions contribute to a negligible amount. So, we display only the 
dominant backgrounds in the table.}.  Similar to the signal 
events, the background processes are also simulated using {\tt Madgraph} and 
then passed to {\tt PYTHIA}. After generating the signal and background events, 
we apply the following kinematic cuts, which are more-or-less in line with those 
applied in a similar analysis by ATLAS collaboration \cite{Aad:2015pfa}. We select the events 
with four isolated muons with $p_{T} > $ 10 GeV and $|\eta|<$ 2.5. The 
isolation criteria imposed on the muons are 
($a$) that the angular separation $\Delta R_{\ell J}$ between the lepton and 
jets\footnote{We reconstruct jets using {\tt FASTJET v3.1.0} \cite{fastjet3} with 
anti-$k_T$ jet algorithm and jet radius R=0.4.} should not be less 
than 0.4, and ($b$) that the sum of the scalar $p_T$ of all 
stable visible particles within a cone of radius $\Delta R = 0.2$ around 
the lepton should not exceed 10~GeV. In Fig.~\ref{minv}, we display 
the $p_T$ distribution of the leading isolated muon. Note that, for the signal events 
the leptons are relatively harder compared to SM backgrounds and this important 
feature can be used as a trigger of such events. For our signal events, 
the muons are coming from the `on-shell' decay of the sneutrino ($\widetilde \nu_1$) 
and thus one can reconstruct $\widetilde \nu_1$ mass using the di-leptonic 
invariant mass. However, for processes like $ZH$, $WH$ with $H \to Z Z^{*}$, one 
$Z$ is on-shell while the other is off-shell, and thus di-muon invariant mass 
will have a long tail with a sharp peak at $M_{Z} \sim$ 91 GeV. Among all 
possible di-muon invariant mass recombinations, the one 
with minimum mass difference $\Delta m = |m_{12} - m_{34}|$ is selected 
where $m_{12}$ and $m_{34}$ are two such di-muon invariant masses. 
We impose a $Z$ veto by 
requiring either of the di-muon invariant masses is greater than 100 GeV. 
Note that, by making such a choice we also reduce the contributions 
coming from processes like associated production of a $Z$ boson 
with $J/\psi$ and/or $\Upsilon$ significantly. In lower 
panel of Fig.~\ref{minv}, the di-muon 
invariant masses are shown for both the signal and background events, 
where for the signal events we show for two representative benchmark 
points BP1 and BP2 with masses $\sim $ 290 GeV and 320 GeV respectively. 
From the figure it is evident that a cut on the di-muon invariant 
mass $m_{\mu\mu} > 100$ GeV would help us to reduce the 
dominant SM backgrounds. 

\begin{table}[h!]
\begin{tabular}{|c|c|c|c|c|}
\hline
Process & $\sigma_{0}$ (fb) & $\sigma_{\rm eff}$ (fb) & $\sigma_{\rm tot.}$ (fb) & Significance \\
\cline{1-5}
\cline{2-5}
$\widetilde\nu_{1}\widetilde\nu^{*}_{1}$  & 4.08 (BP1) &  1.512  &  1.512  &                   \\
                  & 2.64 (BP2)   & 0.98  & 0.98   &            \\ [2mm]
\cline{1-4}
$WH$    & 1380 \cite{higgsXsec}  & 0.014  &  &  8.4 (BP1)           \\ [2mm]
\cline{1-3}
$ZH$   & 868 \cite{higgsXsec}  &  0.0022  & 1.716 & 5.9 (BP2)                  \\ [2mm]
\cline{1-3}
$ZZ$  & 15000 \cite{Baglio:2013tva}  & 1.7  &   &          \\       
\hline
\end{tabular}
\caption{ Event summary for the signal and background events. The 
quantities $\sigma_0$ and $\sigma_{\rm eff}$ represent the production 
cross-section and the effective cross-section respectively. The 
total cross-section is denoted by $\sigma_{\rm tot}$. 
For the $WH$ and $ZH$ processes the Higgs boson is assumed to decay 
to $ZZ^*$ with $Z$ decaying to two muons. We calculate the signal 
significance $\mathcal S = S/\sqrt({S+B})$ with $\mathcal L = 100~{\rm fb}^{-1}$ 
where $S$ and $B$ are total number of signal and background events.}
\label{tab:signi}
\end{table}  
\vspace{0.2cm}
\begin{figure}[h!]%
    {{\includegraphics[width=8cm]
   {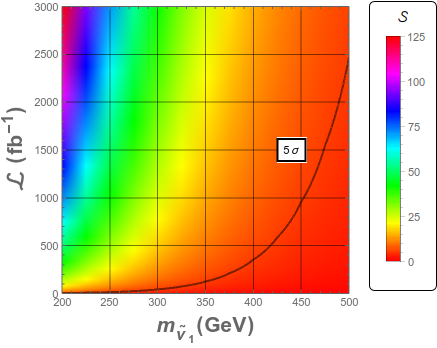} }}%
    \caption{Contour plot in the $\mathcal L - m_{\widetilde \nu_1}$ plane 
for the bino-line neutralino mass parameter of $M_1=300$ GeV. Similar distribution 
can be obtained for $M_1=10$ GeV. }
    \label{sneutrino}%
\end{figure}

In Table~\ref{tab:signi}, we show the production cross-section ($\sigma_0$), 
the effective cross-section ($\sigma_{\rm eff} = \sigma_0 \times \epsilon$, 
with $\epsilon$ being the cut efficiency) for the two benchmark points 
BP1 and BP2 along with the three dominant SM backgrounds $WH$, $ZH$ and $ZZ$ 
with $H \to ZZ^*$. The cross-sections for the signal events are 
calculated using {\tt Madgraph} at the leading order\footnote{ We use 
{\tt Prospino} \cite{Beenakker:1996ed} to calculate the 
$K$-factor associated to the slepton 
pair production process and find $K$=1.2 for slepton masses 
from 200 - 500 GeV.}, while we follow the 
LHC Higgs Cross Section Working Group report \cite{higgsXsec} for the 
$WH$, $ZH$ backgrounds where they are calculated at NNLO QCD and NLO EW. 
The cross-section for $ZZ$ has been taken from 
Ref.~\cite{Baglio:2013tva} calculated at NLO QCD and NLO EW. 
The statistical significance ($\mathcal S$) is 
calculated as $\mathcal S = S/\sqrt({S+B})$, where $S$ and $B$ are 
the total number of signal and background events respectively for 
$\mathcal L = 100~{\rm fb^{-1}}$ luminosity at the 13 TeV run of LHC. From 
the table one can infer that the lightest sneutrinos with masses 
around 300-320 GeV can be discovered using this 4$\mu$ golden channel 
at the early run of 13 TeV LHC. In order to estimate the reach of the 
sneutrinos at the 13 TeV LHC, we now vary the soft-mass parameter 
$(m^2_L)_{33}$ in such a way that the sneutrino mass varies from 
200 GeV to 500 GeV keeping all other parameters same as the BP1 and BP2. 
For each point we again calculate the statistical significance $\mathcal S$ 
as already defined and then vary the luminosity $\mathcal L$. In 
Fig.~\ref{sneutrino}, we display the statistical signal $\mathcal S$ in 
the $\mathcal L - m_{\widetilde \nu_1}$ plane. The black solid line 
indicates the required luminosity for a given sneutrino mass in order 
to obtain 
a 5$\sigma$ statistical significance. We find that using the four muon 
golden channel one can probe the sneutrino masses up to 450 GeV 
with 1000~${\rm fb}^{-1}$ of luminosity at the 13 TeV LHC.

\section{Conclusions}
We revisit the possibility of satisfying anomalous magnetic moment of the muon in the 
paradigm of R-parity violating MSSM. The relevant coupling, $\lambda_{322}$, 
which plays a major role in this process is identified. The 
low energy and neutrino mass constraints 
have been checked and can be rather easily satisfied even at the presence of 
an $\mathcal O$(1) value of this particular R-parity violating coupling. We show that 
this explanation of having large muon $(g-2)$ via R-parity violation can be tested 
directly at the LHC. An artifact of $\mathcal O$(1) $\lambda_{322}$ is 
the decay of the pair produced tau sneutrino in to a final state comprising 
of four muons. This is a so-called ``golden channel" because of large 
signal efficiency and minuscule contribution from the SM backgrounds. 
We analyze all the relevant SM backgrounds and find that sneutrino masses 
upto 450 GeV can be probed with an integrated luminosity of 1000~${\rm fb}^{-1}$ at 
the 13 TeV LHC. Such a channel is yet to be investigated by both the ATLAS and 
CMS collaborations, and it is our hope that this work will motivate them 
to perform a dedicated analysis in this direction in near future.

{\underline {Acknowledgements}} 
\vskip 0.1cm
AC and SC would like to thank the Department of Atomic Energy, Government of 
India for financial support. We gratefully acknowledge Sreerup Raychaudhuri, 
Gautam Bhattacharyya, Sourov Roy, Tuhin S. Roy and Biplob Bhattacherjee for useful 
discussions. SC also thanks Florian Staub for helpful discussions regarding 
{\tt SARAH}. 

\appendix*
\section{}
{
In this appendix we elaborate the SUSY-RPC contribution to
the anomalous magnetic moment of the muon~\cite{Stockinger:2006zn,Martin:2001st,Moroi:1995yh}. 
In general the chargino-sneutrino loop dominates over the neutralino-smuon
loop. We reiterate that when all the mass scales are of the
same order, the chargino-sneutrino loop contribution shown in
Eq.(\ref{eq:a2}) reduces to the form
\begin{widetext}
\begin{eqnarray}
\delta a_{\mu}^{\widetilde\chi^{\pm}}&=&\frac{m_{\mu}}{16\pi^2}
\Bigg\{\frac{m_\mu}{12 M_S^2}\Big(
\cancelto{0}{\frac{m_{\mu}^2}{2M_S^2\cos^2\beta
}}
+g_2^2\Big)+\frac{2 g_2}{3 M_S}\frac{m_{\mu}g_2}{\sqrt{2}M_S\cos\beta}\Bigg\} \nonumber \\
&=&\frac{m_{\mu}^{2} }{192\pi^2 M_s^2}\Bigg\{{g_2^2+\frac{4\sqrt{2}g_2^2}{\cos\beta}}\Bigg\} \nonumber \\
&\simeq&\frac{m_{\mu}^2 g^{2}_2}{192\pi^2 M_s^2}\Bigg\{1+\frac{6}{\cos\beta}\Bigg\}.
\label{chargino-sim}
\end{eqnarray} 
\end{widetext}

In the large $\tan\beta$ limit, Eq.(\ref{chargino-sim}) can be further simplified
to
\begin{eqnarray}
\delta a_{\mu}^{\widetilde\chi^{\pm}}&\simeq&\frac{m_{\mu}^2 g^{2}_2}{192\pi^2 M_s^2}.\frac{6}{\cos\beta}\nonumber \\
&\simeq&\frac{m_{\mu}^2 g^{2}_2}{32\pi^2 M_s^2}\tan\beta.
\end{eqnarray}
 
Similarly, the neutralino-smuon contribution can be written down under
the same approximation as
\begin{eqnarray}
\delta a_{\mu}^{\widetilde\chi^0}&=&\frac{m_{\mu}^2}{192\pi^2 M_s^2}\Big(g_1^2-g_2^2\Big)\tan\beta.
\end{eqnarray}

Therefore, the total RPC-SUSY contribution converges to the form given in Eq.(\ref{ref-apen}).
An interesting point to note that although the one-loop contributions $a_{\mu}^{\widetilde
\chi^{0,\pm}}$ has a term linear in $m_{\widetilde\chi^{0,\pm}}$ (see Eq.(\ref{eq:a1}) and (\ref{eq:a2})) 
but they are not enhanced by $m_{\widetilde\chi^{0,\pm}}$ as compared to the 
other terms~\cite{Stockinger:2006zn}. The reason being, these terms also involve either a factor of 
$y_{\mu}$ or $X_{m1}X_{m2}$, which is again proportional to $(M_{\mu}^2)_{12}$ 
and therefore to $y_{\mu}$. Hence,
all the RPC contributions to $(g-2)_{\mu}$ are of the order of $m_{\mu}^2/M_s^2$
as shown explicitly. On the other hand, for $\mathcal O(1)$ RPV $\lambda$ type
couplings the contribution to $a_{\mu}$ also turns out to be of the same order
as its RPC counterpart.
}





\end{document}